\documentclass{ifacconf}
\usepackage{setspace}
\usepackage{parskip}
\usepackage{graphicx}      
\usepackage{natbib}        
\usepackage{nicefrac}
\usepackage{tikz}
\usepackage{soul}
\usepackage{mdframed}
\usepackage{framed}
\usepackage{amsmath,amssymb}
\usepackage{graphicx}   
\usepackage{comment}
\usepackage{xcolor}
\usepackage{lipsum}
\usepackage{amssymb}
\usepackage[utf8]{inputenc}
\usepackage[utf8]{inputenc}
\usepackage[english]{babel}
\usepackage{enumitem} 
\newcommand{\Rr}{\mathbb{R}}
\newcommand{\mnabla}{\nabla\!}
\theoremstyle{plain}
\newtheorem{dfn}{Definition}
\newtheorem{Thm}{Theorem}

\newtheorem{pro}{Proposition}
\newtheorem{Cor}{Corollary}

\newtheorem{Asu}{Assumption}

\begin{document}
 \pdfoutput=1
\begin{frontmatter}	
\title{ \vspace{-0.9 em} \parbox{1\linewidth}{ \centering Robust trajectory tracking for underactuated mechanical systems without  velocity measurements}}
\author[First]{N. Javanmardi} 
\author[Second]{P. Borja} 
\author[Third]{M. J. Yazdanpanah}
\author[First]{J. M. A. Scherpen}

\address[First]{Jan C. Willems Center for Systems and Control, ENTEG, Faculty of Science and Engineering, University of Groningen, Groningen, The Netherlands (e-mail: {n.javanmardi, j.m.a.scherpen}@rug.nl)}
\address[Second]{School of Engineering, Computing and Mathematics, University of Plymouth, Plymouth, United Kingdom (e-mail:  pablo.borjarosales@plymouth.ac.uk)}
\address[Third]{Control and Intelligent Processing Center of Excellence, School of Electrical and Computer Engineering, University of Tehran, Tehran, Iran (e-mail:  { yazdan}@ut.ac.ir)}
\begin{abstract}
  : In this paper, the notion of contraction is used to solve the trajectory-tracking problem for a class of mechanical systems. Additionally, we propose a dynamic extension to remove velocity measurements from the controller while rejecting matched disturbances. In particular, we propose three control designs stemming from the Interconnection and Damping Assignment Passivity-Based Control approach. The first controller is a tracker that does not require velocity measurements. The second control design solves the trajectory-tracking problem while guaranteeing robustness with respect to matched disturbances. Then, the third approach is a combination of both mentioned controllers. It is shown that all proposed design methods guarantee exponential convergence of the mechanical system to the desired (feasible) trajectory due to the contraction property of the closed-loop system. The applicability of this method is illustrated via the design of a controller for an underactuated mechanical system.            
\end{abstract}
\begin{keyword}
Nonlinear systems, Port-Hamiltonian systems, Trajectory tracking, Contractive system, Disturbance rejection, Underactuated systems, Interconnection and Damping Assignment Passivity-Based Control technique. 
\end{keyword}
\end{frontmatter}
\section{Introduction}
\parskip=1pt
Passivity Based Control (PBC) is a constructive approach to control nonlinear systems, e.g., \cite{ortega2001putting}, including, but not limited to, mechanical systems, e.g., \cite{ortega2017passivity}. Among the PBC techniques, we find the so-called Interconnection and Damping Assignment (IDA), which is often formulated to stabilize systems modeled in the port-Hamiltonian (pH) framework. Notably, pH models are suitable to represent a wide range of physical systems, e.g., \cite{duindam2009modeling}, \cite{van2014port}, while IDA is the most general PBC method to stabilize complex nonlinear systems, e.g., \cite{ortega2004interconnection}. In particular, IDA-PBC has proven suitable to stabilize complex underactuated mechanical systems, e.g., \cite{acosta2005interconnection}.

To solve the trajectory-tracking problem in mechanical systems, it is often useful to find the error dynamics. Then, the controller can be designed such that the error tends to zero, i.e., the trajectory-tracking problem is transformed into a stabilization problem, e.g., \cite{dirksz2009passivity}, \cite{van2020trajectory}. Unfortunately, the traditional definition of the error often yields error dynamics that depend on the original system and the desired trajectories. In \cite{yaghmaei2017trajectory}, the authors provide an alternative to overcome this issue by combining the notion of contractive systems and IDA-PBC to solve the trajectory-tracking problem, originating the so-called timed IDA-PBC approach. In contrast to other methodologies, timed IDA-PBC does not rely on finding the error dynamics and hence, does not use the generalized canonical transformation reported in papers like \cite{fujimoto2001canonical} and \cite{fujimoto2003trajectory}.
An application of the timed IDA-PBC approach  has been recently proposed in \cite{javanmardi2020spacecraft}. Furthermore, in the recent paper \cite{reyes2020family}, the trajectory-tracking problem of flexible-joint robots in the pH framework is addressed by a family of virtual-contraction-based controllers. 

From an application viewpoint, some practical requirements, such as eliminating velocity sensors to reduce costs or noisy estimations and guaranteeing robustness in the presence of external forces or an input measurement bias, may be necessary for the closed-loop system.
Some studies, e.g., \cite{romero2014globally} and \cite{yaghmaei2019output}, propose controllers without velocity measurements by using observers for a class of mechanical systems and pH systems, respectively. However, applying the observer dynamics may increase the complexity of the stability proof. Therefore, in some other publications, for instance, \cite{dirksz2009passivity}, \cite{dirksz2008interconnection},  and \cite{dirksz2012tracking}, dynamic extensions are employed in the pH framework to eliminate the velocity terms from the controller. In \cite{dirksz2008interconnection}, the IDA-PBC method for pH systems with a dynamic extension is investigated to asymptotically stabilize a class of mechanical systems without velocity measurements. In the same line, \cite{dirksz2012tracking} and \cite{dirksz2009passivity} investigate the trajectory tracking approach with only position measurements realized by applying canonical transformation theory, which is reported in \cite{fujimoto2003trajectory} for fully actuated mechanical systems. Furthermore, a saturated PBC scheme not requiring velocity measurements is proposed to solve the tracking problem for robotic arms in \cite{van2020trajectory}. These studies do not guarantee exponential convergence for underactuated systems.
On the other hand, the robustness of the energy-shaping controllers in the presence of external disturbances has been discussed in several references. For instance, robustification by adding an integral action on the passive outputs in a class of pH systems is investigated in \cite{donaire2009addition}. The authors of \cite{romero2013robust} follow a similar idea to design a controller for fully actuated systems with constant disturbances (matched and unmatched). The constant disturbance rejection problem for underactuated mechanical systems is addressed by adding an outer-loop controller to the IDA-PBC in \cite{donaire2017robust}. The mentioned studies focus on the regulation problem and using coordinate transformations to preserve the closed-loop pH form.
In comparison with the mentioned studies, the present work represents a non-trivial extension of the method reported in \cite{yaghmaei2017trajectory}, where velocities measurements are removed from the controller via a dynamic extension.
In contrast to \cite{dirksz2008interconnection} and \cite{donaire2017robust}, which only investigate the regulation problem, the current work elaborates on trajectory tracking.
In  \cite{dirksz2009passivity}, \cite{dirksz2012tracking}, \cite{van2020trajectory} and \cite{romero2013robust}, the given approach is only for fully actuated systems and derived by a transformation obtained by solving a PDE equation. Besides, compared with all mentioned works, the proposed technique simultaneously covers robustification and no-velocity measurement.
In this paper, we focus on addressing the trajectory-tracking problem for a class of mechanical systems, such that the controller rejects constant disturbances and does not need velocity measurements.
Accordingly, the main contributions of the paper are summarized as follows:
\begin{itemize}
	\item   
	We develop a control method to solve the trajectory-tracking problem without velocity terms for a class of underactuated mechanical systems.
	
	\item We propose a robust tracking method that does not require any change of coordinates for a class of underactuated mechanical systems. 
	
	\item  We establish some conditions to combine the two methods mentioned above for a class of underactuated mechanical systems.  
\end{itemize}

The controllers developed in this work are based on contraction and dynamic extensions. We stress that
the convergence property of a contractive system guarantees that all the trajectories of the system converge exponentially  as $t \to \infty$. Therefore, all the tracking methods proposed in this paper ensure exponential convergence to the desired trajectory.  

The rest of the paper is organized as follows. Section \ref{sec:ConpH} briefly recalls a class of contractive pH systems. The class of mechanical pH systems under study is introduced in Section \ref{sec:Mec moddel}. We propose a tracking controller depending only on position terms in Section \ref{sec:no-velocity}. A robust tracking method is developed in Section \ref{sec:robust-track}. Section \ref{sec:no-velocity-robust} contains a robust technique to track exponentially a reference trajectory without velocity measurements. The performance of the proposed method  is simulated in Section \ref{sec:sim} for an underactuated mechanical system. Section \ref{sec:con} is devoted to the concluding remarks.\\ 
\textbf{Notation:}
 In the subsequent sections, $A\succ0$ ($A\succeq0$) means that the matrix A is positive definite (positive semi-definite), respectively. $\nabla H$ is defined as $[\,\frac{\partial H}{\partial x_1}, \frac{\partial H}{\partial x_2}, ... \frac{\partial H}{\partial x_n}]\,^\top$ for a continuously differentiable function of $H(x):\mathbb{R}^n \longrightarrow \mathbb{R}_{+} $, and ${\nabla}^2 H$ is a matrix whose $ij$th element is $\frac{{\partial}^2 H}{\partial x_i \partial x_j}$. For a full-rank matrix $g \in \Rr^{n \times m}$, with $m \leq n$, we define $g^{\dagger}\triangleq(g^{\top}g)^{-1}g^{\top}$.
\vspace{-0.3em}
\section{Preliminaries}
\label{sec:Review}
\subsection{Contractive pH systems}
\label{sec:ConpH}
Consider the input-state-output representation of pH systems, which is given by
\begin{equation}
\label{eq:PHOL}
\begin{split}
&\dot{x}=(J(x)-R(x))\nabla H(x) + g(x) u,\\&
y=g^\top(x)\nabla H(x),\quad u,y \in \Rr^m,
\end{split}
\end{equation}
where $D_0$ is the state space of the system, which is an open subset of $\Rr^n$, $x(t)\in D_0\subset \Rr^n$ is the state, the interconnection matrix $J: D_0 \to \Rr^{n\times n}$ is skew-symmetric, the damping matrix  $R:  D_0 \to \Rr^{n\times n} $ is positive semi-definite, $H: D_0 \to \Rr_{+}$ is the system's Hamiltonian, the input matrix $g: D_0 \to \Rr^{n \times m}$ satisfies $\mbox{rank}(g)=m \leq n$, $u,\ y$ are the input vector and the output vector, respectively. To simplify the notation in the rest of the paper, we define the matrix $F:D_0 \to \Rr^{n \times n}$, $F(x)\triangleq J(x)-R(x)$.

In this paper, we tackle the trajectory-tracking problem. To this end, we exploit the properties of \textit{contractive} systems, particularly, the property that all the trajectories of a contractive system converge exponentially to each other as $t\to \infty$.
Therefore, the control problem is reduced to finding a controller such that the closed-loop system is contractive, and the desired trajectory is a feasible trajectory. Before proposing the control approach, it is necessary to introduce the following definition.

\begin{dfn}
	\label{definition1}
	Consider the system \eqref{eq:PHOL} and let $\mathbb{T}$ be an open subset of $\Rr_{+}$. Then,
	$x^\star(t):\mathbb{T}\rightarrow\Rr^n$ is said to be a \textit{feasible trajectory} if there exists $u^\star(t):\mathbb{T}\rightarrow \Rr^m$ such that, for all $t\in \mathbb{T}$, the following relation holds:
	\begin{equation*}
	\dot{x}^\star=F(x^\star(t))\nabla H(x^\star(t))+g(x^\star(t))u^\star(t).
	\end{equation*}
\end{dfn}

The following theorem establishes the main results on contractive systems, which are the cornerstone of the results presented in Sections \ref{sec:no-velocity}, \ref{sec:robust-track}, and \ref{sec:no-velocity-robust}. For further details on timed IDA-PBC and the proof of Theorem \ref{Th1}, we refer the reader to \cite{yaghmaei2017trajectory}.
 \begin{Thm}[\cite{yaghmaei2017trajectory}]
 	\label{Th1}
 	Consider the following system
 	\begin{equation}
 	\label{closed sys}
 	\dot{x}=F_d \mnabla H_d(x, t),
 	\end{equation}
 	with $F_d\triangleq J_d-R_d$, where $J_d=-J^\top_d$ and $R_d=R^\top_d\succeq 0$ are the (constant) desired interconnection and damping matrices, respectively.  The system \eqref{closed sys} is contractive on the open subset $D_0 \subseteq \Rr^n $ if:
 	\begin{enumerate}[label=(\roman*)]
 		\item  All the eigenvalues of $F_d$ have strictly negative real parts.
 		\item  The desired Hamiltonian function $H_d(x,t):\Rr^n \times \Rr_+  \to \Rr_{+}$ satisfies
 		\begin{equation}
 		\label{cond3}
 		\begin{split}
 		&\alpha I\prec \nabla^2 H_d(x, t) \prec \beta I,\quad \forall x \in D_0,
 		\end{split}
 		\end{equation}
 		for constants $\alpha,\beta$, such that $0<\alpha<\beta$.
 		\item There exists a positive constant $\varepsilon$ such that 
 		\begin{equation}
 		\label{N}
 		N \triangleq \begin{bmatrix}
 		F_d & \left( 1-\frac{\alpha}{\beta} \right) F_dF_d^\top\\ -\left(1-\frac{\alpha}{\beta}+\varepsilon\right)I & -F_d^\top
 		\end{bmatrix},
 		\end{equation}
 		has no eigenvalues on the imaginary axis. $\blacksquare$
 	\end{enumerate}
 \end{Thm} 
 \vspace{-0.2em}
\begin{rem}
		\label{rem:contr-struc}
		The proof of Theorem \ref{Th1} only requires $F_d$ to be Hurwitz (see \cite{yaghmaei2017trajectory}). Hence, the condition $F_d+F^\top_d \preceq 0$ (or $Rd \succeq 0$) is not necessary. However, if such a condition is not satisfied, then system \eqref{closed sys} has not a pH structure. Using this fact is precisely the property that admits developing some control methodes without a coordinate transformation in the next sections.		
\end{rem} 
\subsection{Mechanical systems in the pH framework}
\label{sec:Mec moddel}
We restrict our attention to mechanical systems, without natural dissipation, influenced by matched constant disturbances. Such systems admit a pH representation of the form
\begin{equation}
\label{open-loop mech}
\arraycolsep=1pt
\def\arraystretch{1.2}
\begin{array}{cl}
&\begin{bmatrix}
\dot{q} \\ \dot{p}
\end{bmatrix}= \begin{bmatrix}
0 & \quad  I \\ -I & \quad 0
\end{bmatrix}\begin{bmatrix}
{{\nabla _{{q}}}{H(q,p)}} \\ 
{{\nabla _{{p}}}{H(q,p)}}
\end{bmatrix} + \begin{bmatrix}
0 \\ G(q)
\end{bmatrix}(u(t)+d),\\&
H(q,p)=\displaystyle\frac{1}{2}{p^{\top}}M^{-1}(q)p+V(q), 
\end{array}   
\end{equation}
where $q,p\in {\mathbb{R}^{n}}$ denote the generalized positions and momenta, respectively, $u\in\Rr^{m}$ is the input vector, $d\in {\mathbb{R}^{m}}$ is the constant disturbance, $M:\Rr^{n}\to\Rr^{n \times n}$ is the inertia matrix, which is \textit{positive definite}, $V:\Rr^{n}\to\mathbb{R}_{+} $ is the potential energy of the system, and $G:\mathbb{R}^{n} \to \mathbb{R}^{n\times m}$ is the input matrix, satisfying $\mbox{rank}(G)=m \leq n$.
\section{Tracking method without velocity measurements} 
\label{sec:no-velocity}  
This section presents a controller that solves the trajectory-tracking problem without velocity measurements for mechanical systems. To this end, we propose a dynamic extension to inject damping without measuring the velocities of the system. Then, we apply the results of Theorem \ref{Th1} to propose a contractive closed-loop dynamics, which ensures that the trajectories of the closed-loop system converge to the desired ones.
\begin{Asu}
  \label{ass:M}
		The inertia matrix $M \in \mathbb{R}^{n \times n}$ is constant.	
\end{Asu}

Consider the following closed-loop dynamics
\begin{equation}
\label{Eq:target1}
\begin{bmatrix}
{{\dot{q}}} \\[.05cm]
{{\dot{p}}}\\[.05cm]
{{\dot{x}_e}}
\end{bmatrix}=\begin{bmatrix}
0 & J_{d_{12}} & 0\\[.05cm]
-J_{d_{12}}^{\top} & 0 & {S}_1\\[.05cm]
{S}_2 & 0 & F_e                        
\end{bmatrix}\begin{bmatrix}
{{\nabla_{{q}}}{{H}_{d_1}(q,p,x_e,t)}}\\[.05cm]
{{\nabla_{{p}}}{{H}_{d_1}(q,p,x_e,t)}}\\[.05cm]
{{\nabla_{{x_e}}}{{H}_{d_1}(q,p,x_e,t)}}                 
\end{bmatrix},
\end{equation} 
\begin{equation}
\arraycolsep=1pt
\def\arraystretch{1.2}
\begin{array}{rcl}
\label{Eq:H-target1}
{H}_{d_1}(q,p,x_e,t)&=&\frac{1}{2}p^\top M_d^{-1}p+\frac{1}{2}p_e^\top M_e^{-1}p_e + V_{d_1}(q,q_e,t),
\end{array}
\end{equation} 
where $x_e(t)=[q_e^{\top}(t), p_e^{\top}(t)]^{\top} \in {\mathbb{R}^{2m}}$, with $m \leq n$, denotes the state of the controller, $H_{d_1}:\mathbb{R}^{2n} \times \mathbb{R}^{2m} \times \mathbb{R}_{+} \to \mathbb{R}_{+} $ is the closed-loop energy function, $V_{d_1} :\mathbb{R}^{n} \times \mathbb{R}^{m} \times \mathbb{R}_{+} \to \mathbb{R}_{+}$ is the desired potential energy function,
  the desired  constant inertia matrix $M_d \in \mathbb{R}^{n \times n}$ and the  constant  controller inertia matrix $M_{e} \in \mathbb{R}^{m \times m}$ are positive definite. Moreover, the constant matrix $F_e\in\mathbb{R}^{2m\times 2m}$ is given by
$F_e \triangleq J_e-R_e $, where $J_e=-J^{\top}_e$ and $R_e=R^{\top}_e\succ 0$ are the constant controller interconnection and damping matrices, respectively. Furthermore, $ S_1 \in {\mathbb{R}^{n\times 2m}}$, and $S_2 \in {\mathbb{R}^{2m\times n}}$ are full-rank matrices. In particular, $S_1$ has the following structure	
\begin{equation}
\label{Eq:S_1}
S_1= \left[ {\begin{array}{@{}cc@{}}
	s_{11},& s_{12}
	\end{array}} \right],
\end{equation}
where $s_{11}\in \mathbb{R}^{n\times m} $ and $s_{12}\in \mathbb{R}^{n\times m} $. Then, from Theorem \ref{Th1} and Remark \ref{rem:contr-struc}, we have the following corollary for the closed-loop system \eqref{Eq:target1}.
\begin{Cor}
	\label{Cor:contract-w}
	Consider the closed-loop dynamics \eqref{Eq:target1}. Suppose that:   
	\begin{itemize}
		\item [\textbf{C1.}] The matrix
		\begin{equation}
		\label{Eq:P_1}
		P_{1}\triangleq  \left[ {\begin{array}{@{}ccc@{}}
			0 & J_{d_{12}} & 0\\[.05cm]
			-J^{\top}_{d_{12}} &0&{S}_1\\[.05cm]
			{S}_2&0&F_e
			\end{array}} \right]
		\end{equation}	
		is Hurwitz.
		\item [\textbf{C2.}] There exist constants $0<\alpha_1< \beta_1$ such that $ H_{d_1}(\xi,t)$ in \eqref{Eq:H-target1}, with $\xi=[q^\top, p^{\top}, x_e^\top]^\top$, satisfies the inequality
		\begin{equation}
		\label{Eq:bondcond1}
		\alpha_1 I\prec \nabla_{\xi}^2 H_{d_1}(\xi,t) \prec \beta_1 I,\; \forall \xi \in D_1\subseteq \Rr^{2n+2m},
		\end{equation}
	\end{itemize}
 
	The system $\eqref{Eq:target1}$ is contractive on $D_1$ if there exists a positive constant $\varepsilon_1$ such that 
	\begin{equation}
	\label{Eq:N1}
	N_1 = \begin{bmatrix}
	P_{1} & \left(  1-\frac{\alpha_1}{\beta_1} \right) P_{1}P_{1}^\top\\ -( 1-\frac{\alpha_1}{\beta_1}+\varepsilon_1)I & -P_{1}^\top
	\end{bmatrix},
	\end{equation}
has no eigenvalues on the imaginary axis. $\square$
\end{Cor}	 
\begin{rem}
	The first two terms in the energy function \eqref{Eq:H-target1}, correspond to the desired kinetic energy of the plant and the kinetic energy of the controller, respectively. Whereas $V_{d_1}(q,q_e,t)$ is the potential energy function that couples the plant with the controller.
\end{rem}
\begin{rem}
Note that the symmetric part of $P_1$ has no definite sign. However, an appropriate selection of $J_{d_{12}}$, $S_1$, $S_2$, and $F_e$, may guarantee that $P_1$ is Hurwitz. Moreover, the closed-loop structure can be achieved with a control law that does not include velocity terms or coordinate transformations. 	
\end{rem}
\vspace{-0.5 mm}

The next theorem provides a controller for mechanical systems that achieves trajectory tracking without measuring velocities.
\begin{Thm}
 	\label{Pro:UT-w}
 	Consider the mechanical system \eqref{open-loop mech}  under Assumption \ref{ass:M}, with $d=0$, and the feasible desired trajectory $x^\star(t)=[{q^\star}^\top,{p^\star}^\top]^\top$. 
 	If there exist $J_{d_{12}}$, $M_{d}$, $M_e$, $V_{d_1}$, ${S}_1$, ${S}_2$, and $F_e$ such that:	
 \begin{enumerate}[label=(\roman*)]
 		\item   The following matching equations hold\footnote{We omit the arguments to ease the readability.}
\begin{eqnarray}
&&\hspace{-1.2 cm}M^{-1}p=J_{d_{12}}M_d^{-1}p, \label{Eq:matcing1-p1}\\&
&\hspace{-1.2 cm}G^\perp\bigg(\nabla_q V-J_{d_{12}}^{\top}\nabla_q V_{d_1}+s_{11}\nabla_{q_e} V_{d_1}+s_{12}M_e^{-1}p_e\bigg)=0.\qquad\label{Eq:matcing1-p3}
\end{eqnarray}

 		\item The conditions given in Corollary \ref{Cor:contract-w} are satisfied. 	
 		\item The following equation holds
 		\begin{equation*}
 		\label{Eq:refer1}
 		\dot\xi^{\star} = P_{1}{\nabla_\xi}{{H_{d_1}}({\xi^\star, t})},
 		\end{equation*} 
 		where ${\xi^\star} \triangleq [{x^\star}^\top\!, {x_e^\star}^\top\!]^\top$ and ${p^\star}= M\dot q^\star$.
 	\end{enumerate}
  
 	Then, the input signal
\begin{equation}
 	\label{Eq:UT-w}
 	\begin{array}{rcl}
 	u &=& G^ \dagger(q)\left(-J^{\top}_{d_{12}}\nabla_{q} {{V_{d_1}}(q,q_e,t)}+{S}_1\mnabla_{x_e}{{H_{d_1}}(\xi,t)}\right.\\[0.1cm] &&+\left.
 	\nabla_q V(q)\right),
 	\end{array}
 	\end{equation}
 	with
 	\begin{equation}
 	\label{Eq:extension1}
 	\dot{x}_e={S}_2\nabla_{q}{{V}_{d_1}}(q,q_e,t)+F_e\nabla_{x_e} {{{H}_{d_1}}(\xi,t)},
 	\end{equation}
guarantees that the trajectories of the closed-loop system converge exponentially to $x^\star(t)$.
 		\begin{pf} 
 			\normalfont 
 Because of (i), \eqref{open-loop mech} in closed-loop with \eqref{Eq:UT-w} and \eqref{Eq:extension1} yields \eqref{Eq:target1}. Moreover, from (ii),
           the closed-loop system is contractive.
 Furthermore, (iii) ensures that $\xi^\star(t)$ is a trajectory of \eqref{Eq:target1}. Accordingly, due to the convergence property of contractive systems \cite[Theorem 1]{lohmiller1998contraction}, the trajectories of the closed-loop system  \eqref{Eq:target1} converge exponentially to $\xi^\star(t)$.
 $\blacksquare$
 		\end{pf} 
\end{Thm}

\subsection{Parameter design for solving the matching PDEs.}
	\label{se:Par-cte}
	Consider \eqref{open-loop mech} with constant input matrix $G$, constant inertia matrix $M$, and without external disturbance, i.e., $d=0$. Choose $M_d$ according to one of the following cases:
	\begin{itemize}
        \item [(a)] $J_{d_{12}}=M^{-1}M_{d}$ in the case of total energy shaping.
		\item [(b)] $J_{d_{12}}=I$,  $M_d\!=\!M$ in the case of potential energy shaping. 
	\end{itemize}
 
	Consider the following target system:
	\begin{equation}
	\label{Eq:targetIDA}
	\begin{bmatrix}
	{{\dot{q}}} \\[.05cm]
	{{\dot{p}}}
	\end{bmatrix}=\begin{bmatrix}
	0 & J_{d_{12}}\\[.05cm]
	-J_{d_{12}}^{\top} & 0                        
	\end{bmatrix}\begin{bmatrix}
	{{\nabla_{{q}}}{{H}_{d}(q,p)}}\\[.05cm]
	{{\nabla_{{p}}}{{H}_{d}(q,p)}}                 
	\end{bmatrix},
	\end{equation} 
	\begin{equation}
	\arraycolsep=1pt
	\def\arraystretch{1.2}
	\begin{array}{rcl}
	\label{Eq:H-target}
	{H}_{d}(q,p)&=&\frac{1}{2}p^\top M_d^{-1}p+\tilde{V}_{d}(q),
	\end{array}
	\end{equation} 
	where $\tilde V_d(q)$ is the desired potential energy.
	Hence, to solve the conventional IDA-PBC problem, the following matching equation needs to be solved
	\begin{eqnarray}
	\label{solvability22}
	&G^\perp\Big(\nabla_q V(q)-J_{d_{12}}^{\top}\nabla_q \tilde V_d(q)\Big)=0.
	\end{eqnarray}
	
	Now, consider the tracking approach presented in Theorem \ref{Pro:UT-w}.
	Set $S_1= G[k_{11},k_{12}]$, where $k_{11},k_{12} \in \mathbb{R}^{m \times m}$. Accordingly, we have $s_{11}=Gk_{11}$ and $s_{12}=Gk_{12}$ in \eqref{Eq:S_1}. Therefore, 
 the matching equation \eqref{Eq:matcing1-p3} is reduced to

	\begin{eqnarray}
&G^\perp\Big(\nabla_q V(q)-J_{d_{12}}^{\top}\nabla_q V_{d_1}(q,q_e,t)\Big)=0 	\label{solvability11}.	    
	\end{eqnarray} 

	The set of solutions to \eqref{solvability11} is the same as the set of solutions to \eqref{solvability22}. Therefore, if the general solution to \eqref{solvability22} is available, one can use it to solve \eqref{solvability11}.
\section{Robust tracking method subject to matched disturbances}
\label{sec:robust-track}
In this section, the main objective is to design a dynamical controller
such that the mechanical system \eqref{open-loop mech} exponentially tracks the desired signal in the presence of constant matched disturbances. 
To this end, we design a contractive closed-loop system. In particular, we propose a dynamical controller $u(t)=v(x(t),\zeta(t))$, where $\zeta(t) \in  {\mathbb{R}^{m}} $, that rejects the unknown disturbance and guarantees the contraction property for the closed-loop system.
	\begin{Asu}
 \label{ass:G}
		The input matrix $G$ is constant.	
	\end{Asu}
 
	Consider the following closed-loop system subject to a matched disturbance $d\in\mathbb{R}^m$:
 \begin{equation}
 \label{Eq:target2-d}
 \arraycolsep=1pt
\def\arraystretch{1}
\begin{array}{cl}
&\begin{bmatrix}
{{\dot{q}}} \\[.05cm]
{{\dot{p}}}\\[.05cm]
{{\dot{\zeta}}}
\end{bmatrix}=\begin{bmatrix}
0 & {J}_{d_{12}} & 0\\[.05cm]
-{{J}^{\top}_{d_{12}}} & -{R}_d & {W}_1\\[.05cm]
{W}_2 & {W}_3 & 0                       
\end{bmatrix}\begin{bmatrix}
{{\nabla_{{q}}}{{H}_{d_2}(q,p,\zeta,t)}}\\[.05cm]
{{\nabla_{{p}}}{{H}_{d_2}(q,p,\zeta,t)}}\\[.05cm]
{{\nabla_{{\zeta}}}{{H}_{d_2}(q,p,\zeta,t)}} \end{bmatrix}+ \begin{bmatrix}
{0}\\[.05cm]
{G}\\[.05cm]
{0}                  
\end{bmatrix}d,\\&
{H}_{d_2}(q,p,\zeta,t)=\frac{1}{2}p^\top {M}_d^{-1}(q)p+V_{d_2}(q,t)
\\&\hspace{20mm}+\frac{1}{2}(\zeta(t)-\gamma_1(t))^{\top}K_{\zeta}(\zeta(t)-\gamma_1(t)),
\end{array} 
\end{equation}
where $\zeta(t)\in {\mathbb{R}^{m}}$, with $m \leq n$, is the state of the controller; ${H}_{d_2}:\mathbb{R}^{2n} \times \mathbb{R}^{m} \times \mathbb{R}_{+} \to \mathbb{R}_{+}$ represents the closed-loop energy function; $V_{d_2}:\mathbb{R}^{n} \times \mathbb{R}_{+} \to \mathbb{R}_{+}$ is the desired
potential energy function, $\gamma_1(t)\in {\mathbb{R}^{m}}$; $K_\zeta \in {\mathbb{R}^{m \times m}}$ is positive definite. Moreover,  the desired inertia matrix $M_d: \mathbb{R}^{n} \to  \mathbb{R}^{n \times n}$ and the damping matrix ${R}_d \in\mathbb{R}^{n\times n}$ are positive definite, $W_1 \in {\mathbb{R}^{n\times m}}$ and $W_2, W_3 \in {\mathbb{R}^{m\times n}}$  satisfy $\mbox{rank}(W_1)=\mbox{rank}(W_2)=\mbox{rank}(W_3)=m \leq n$.

In order to show that \eqref{Eq:target2-d} is contractive, we define the following energy function
\begin{equation}
\arraycolsep=1pt
\def\arraystretch{1.2}
\begin{array}{rcl}
\label{Eq:hamiltonian-target2}
&\bar {H}_{d_2}(q,p,\zeta,t)\triangleq\frac{1}{2}p^\top M_d^{-1}(q)p+V_{d_2}(q,t)
\\&\quad+\frac{1}{2}(\zeta(t)-\gamma_1(t)+\mu_1)^{\top}K_{\zeta}(\zeta(t)-\gamma_1(t)+\mu_1),
\end{array}
\end{equation} 
with $\mu_1 \triangleq  (W_1 K_{\zeta})^\dagger  G d$. Then rewrite the closed-loop system \eqref{Eq:target2-d} as 
\begin{equation} 
\label{Eq:target2}
\begin{bmatrix}
{{\dot{q}}} \\[.05cm]
{{\dot{p}}}\\[.05cm]
{{\dot{\zeta}}}
\end{bmatrix}=\begin{bmatrix}
0 &  J_{d_{12}} & 0\\[.05cm]
- J_{d_{12}}^{\top} & -{R}_d & {W}_1\\[.05cm]
{W}_2 & {W}_3 & 0                        
\end{bmatrix}\begin{bmatrix}
{{\nabla_{{q}}}{\bar{H}_{d_2}(q,p,\zeta,t)}}\\[.05cm]
{{\nabla_{{p}}}{\bar{H}_{d_2}(q,p,\zeta,t)}}\\[.05cm]
{{\nabla_{{\zeta}}}{\bar{H}_{d_2}(q,p,\zeta,t)}}                 
\end{bmatrix}.
\end{equation}

The next corollary follows from Theorem \ref{Th1}. 

\begin{Cor}
	\label{Cor:contract-r}
	The system $\eqref{Eq:target2}$ is contractive on an open subset $D_2 \subseteq \Rr^{2n+m}$ if:
	\begin{itemize}
		\item [\textbf{C1.}] The matrix
		\begin{equation}
		\label{Eq:P_2}
		P_2 \triangleq\begin{bmatrix}
		0 &  J_{d_{12}} & 0\\[.05cm]
		- J_{d_{12}}^{\top} & -{R}_d & {W}_1\\[.05cm]
		{W}_2 & {W}_3 & 0                       
		\end{bmatrix},
		\end{equation}
		is Hurwitz.
		\item [\textbf{C2.}] The following condition is satisfied for the positive constants $\alpha_2,\beta_2$ (with $\alpha_2<\beta_2$):
		\begin{equation}
		\label{Eq:bondcond2}
		\begin{split}
		&\alpha_2 I\prec \nabla_{\eta}^2 \bar H_{d_2}(\eta,t) \prec \beta_2 I,\quad \forall \eta \in {D}_2, 
		\end{split}
		\end{equation}
		where $\eta=[q^\top, p^\top ,\zeta^\top]^\top$.
		\item [\textbf{C3.}] The following matrix has no eigenvalues on the  imaginary axis for a positive constant $\varepsilon_2$:	
		\begin{equation}
		\label{Eq:N2}
		N_2 = \begin{bmatrix}
		P_2 &  \big(1-\frac{\alpha_2}{\beta_2}\big) P_2P^\top_2\\ -( 1-\frac{\alpha_2}{\beta_2}+\varepsilon_2)I & -P^\top_2
		\end{bmatrix}.\quad \square
		\end{equation} 
	\end{itemize}
\end{Cor}

The next theorem provides a dynamical controller that solves the trajectory-tracking problem  under the effect of constant matched disturbances for mechanical systems. To simplify the notation in the rest of the paper, we define $\Theta(q,p,t)\triangleq\frac{1}{2}\nabla_q \big({p}^\top M_d^{-1}(q)p\big)+\nabla_q  V_{d_2}(q,t)$. 

\begin{Thm}
	\label{Pro:UT-r}
	Consider the mechanical system \eqref{open-loop mech}, with a feasible trajectory $x^\star(t)\!=\![{q^\star}^\top\!\!,{p^\star}^\top\!]^\top$, and a constant input matrix $G$.
    Assume that there exist $J_{d_{12}}$, $M_d$, $V_{d_2}$,  ${R}_d, K_\zeta$ and $W_i$, for $i\kern-.3em\in\kern-.3em \{1,2,3\}$, such that: 
	 \begin{enumerate}[label=(\roman*)]
		\item  The following matching equations hold
		\begin{eqnarray}
        &&\hspace{-1.2 cm}M^{-1}(q)p=J_{d_{12}}M_d^{-1}(q)p,\label{Eq:matcing2-p1}\\
		&&\hspace{-1.2 cm}G^\perp\Big(\nabla_q (p^\top M^{-1}(q)p)-J^\top_{d_{12}}\nabla_q (p^\top M_d^{-1}(q)p) \nonumber\\
		&&\hspace{-1.2 cm}-2{R}_d  M_d^{-1}(q)p\Big)=0,
		\label{Eq:matcing2-p2}\\
		&&\hspace{-1.2 cm}G^\perp\Big(\nabla_q V\!-J^\top_{d_{12}}\nabla_q V_{d_2}+ W_1K_{\zeta}(\zeta-\gamma_1)\Big)\!=\!0.\label{Eq:matcing2-p3}\end{eqnarray}
		\item The conditions of Corollary \ref{Cor:contract-r} are 
		satisfied. 		
		\item The following equation is satisfied:
		\begin{equation*}
		\label{Eq:refer2}
		{W}_2\Theta(q^\star,p^\star,t)+{W}_3 M_d^{-1}(q^\star)p^\star=0,
		\end{equation*}
		where ${p^\star}= M(q^\star)\dot q^\star$.			
	\end{enumerate} The controller
	\begin{equation}
	\label{Eq:UT-R}
	\begin{split}
	u = G^\dagger\Big( -J^\top_{d_{12}}\Theta(q,p,t)-{R}_d M_d^{-1}(q)p &\\+W_1 K_\zeta (\zeta(t)-\gamma_1(t))
	+\mnabla_q H(q,p)\Big), &
	\end{split}
	\end{equation}
with
	\vspace{-0.3cm}
\begin{eqnarray}
 \gamma_1(t)&=&(G^\top{W}_1K_\zeta)^{-1}\big(G^\top(-J^\top_{d_{12}}\Theta(q^\star,p^\star,t) \label{Eq:gamma1} \nonumber \\&-&{R}_d  M_d^{-1}(q^\star)p^\star)-G^\top\dot{p}^\star)\big), \\\dot \zeta(t)&=&{W}_2 \Theta(q,p,t
 )+{W}_3 M_d^{-1}(q)p, \label{Eq:zeta}
\end{eqnarray}
	makes the system \eqref{open-loop mech} a local exponential tracker for $x^\star(t)$, while eliminating the effect of the constant disturbance $d$.	
\begin{pf}
 Set the controller of the system \eqref{open-loop mech} as \eqref{Eq:UT-R}. Hence, from (i), the closed-loop is given by \eqref{Eq:target2}, and because of (ii), it is contractive. Moreover, the desired signals $(x^\star(t),\zeta^\star)$, where $\zeta^\star \in {\mathbb{R}^{m}}$ is constant, are evaluated in \eqref{Eq:target2} as follows:
		\begin{align}				
		&\dot p^\star=-J^\top_{d_{12}}\nabla_q \bar H_{d_2}(\eta^\star,t)-{R}_d  M_d^{-1}(q^\star)p^\star\nonumber\\&+{W}_1 K_\zeta (\zeta^\star-\gamma_1(t)+\mu_1) \label{Eq:refer2-x},\\&
        0={W}_2\nabla_q {\bar H_{d_2}(\eta^\star,t)}+{W}_3 M_d^{-1}(q^\star)p^\star.\label{Eq:refer2-zeta}
		\end{align}
		
		Multiply both sides of \eqref{Eq:refer2-x} by the invertible matrix $[G^\perp,G^\top]^\top$ and replace 
        $\small \nabla_q \bar H_{d_2}(\eta^\star,t)$ by $\Theta(q^\star,p^\star,t)$. From (iii) and \eqref{Eq:gamma1}, it follows that \eqref{Eq:refer2-x} and \eqref{Eq:refer2-zeta} are satisfied with $\zeta^\star = -\mu_1 $. Hence, due to the convergence property of contractive systems, all the trajectories exponentially converge to the desired ones. Note that there is no disturbance information in the controller. Besides, since $\zeta^\star = -\mu_1$, the effect of the disturbance in \eqref{open-loop mech} is eliminated by the controller as the time tends to infinity.
        $\blacksquare$
	\end{pf}
\end{Thm}

\begin{rem} 
	\label{solvability2}
	Suppose Assumption \ref{ass:M}. Choosing ${W}_1=GK_2$ and ${R}_d=GK_vG^\top$, where $K_2 \in \Rr^{m \times m}$ and $K_v \in \Rr^{m \times m}$, the matching equations \eqref{Eq:matcing2-p2} and \eqref{Eq:matcing2-p3} reduce to
	\begin{equation}
	\label{solvability12}
	\begin{array}{cl}	&G^\perp\Big(\nabla_q V-J^\top_{d_{12}}\nabla_q  V_{d_2}(q,t)\Big)=0.
	\end{array}
	\end{equation}
The solution to \eqref{solvability12} can be achieved by solving the matching equation for regulation via IDA-PBC, given in \eqref{solvability22}.   	
\end{rem}
\section{Robust tracking method without velocity measurements}
\vspace{-0.1cm}
\label{sec:no-velocity-robust}
Motivated by the approaches provided in Sections \ref{sec:no-velocity} and \ref{sec:robust-track}, a robust tracking method depending only on position terms is proposed for mechanical systems in this section. To this end, a contractive closed-loop system is designed. Then, we propose a dynamical controller with no velocity measurements to track the desired trajectory while rejecting the constant disturbance.

Consider the following closed-loop system subject to a matched disturbance $d$:
\begin{equation}
\begin{array}{cl}
&\begin{bmatrix}
{{\dot{q}}} \\[.05cm]
{{\dot{p}}}\\[.05cm]
{{\mathcal{\dot Z}}}\\[.05cm]
\end{bmatrix}=\begin{bmatrix}
0 & J_{d_{12}} & 0\\[.05cm]
-J_{d_{12}}^{\top} & 0&\digamma_1\\[.05cm]
\digamma_2 & 0 & \digamma_3\\[.05cm]               \end{bmatrix}\begin{bmatrix}
{{\nabla_{{q}}}{{H}_{d_3}(q,p,\mathcal{Z},t)}}\\[.05cm]
{{\nabla_{{p}}}{{H}_{d_3}(q,p,\mathcal{Z},t)}}\\[.05cm]
{{\nabla_{\mathcal{Z}}}{{H}_{d_3}(q,p,\mathcal{Z},t)}}\\[.05cm]            \end{bmatrix}+ \begin{bmatrix}
{0}\\[.05cm]
{G}\\[.05cm]
{0}\\[.05cm]                
\end{bmatrix}d,\\&
\label{Eq:hamiltonian3-d}
{H}_{d_3}(q,p,\mathcal{Z},t)=\frac{1}{2}p^\top M_d^{-1}p
+V_{d_3}(q,z_1,t)\\&
+\frac{1}{2}(z_2(t)-\gamma_2(t))^{\top}K_{z}(z_2(t)-\gamma_2(t)),
\end{array}
\end{equation}
 where $\mathcal{Z}(t)\mkern-8mu=[z_1^\top(t),z_2^\top(t)]^\top$ with $z_1, z_2\in\Rr^{m}$, $m\!\leq n$, indicates the states of the controller, 
 $M_d \in \mathbb{R}^{n \times n}$ is  the desired  constant inertia matrix,
 ${H}_{d_3}\mkern-8mu:\mathbb{R}^{2n} \times \mathbb{R}^{2m} \times \mathbb{R}_{+} \to \mathbb{R}_{+}$ is the closed-loop energy function,   $V_{d_3}\mkern-8mu: \mathbb{R}^{n} \times \mathbb{R}^{m} \times \mathbb{R}_{+} \to \mathbb{R}_{+}$ is the desired potential energy function. Moreover, $K_z\mkern-3mu\in\! {\mathbb{R}^{m \times m}}\mkern-3mu\succ\mkern-3mu0$, $\gamma_2(t)\mkern-4mu\in \Rr^m$, and $\digamma_1 \in {\mathbb{R}^{n\times 2m}}$, $\digamma_2 \in {\mathbb{R}^{2m\times n}}$, and $\digamma_3  \in\mathbb{R}^{2m\times 2m}$ are given by	
\begin{equation}
\label{Eq:digamma12}
\digamma_1= \left[ {\begin{array}{@{}cc@{}}
	\Gamma_{11},&\Gamma_{12}
	\end{array}} \right], \digamma^\top_2= \left[ {\begin{array}{@{}cc@{}}
	\Gamma_{21}^\top,& \Gamma_{22}^\top
	\end{array}} \right],
\digamma_3=\begin{bmatrix}
-\Gamma_{33}&0\\0&0\end{bmatrix} \succeq  0,
\end{equation}
where $\Gamma_{11}, \Gamma_{12} \in \mathbb{R}^{n\times m} $ and $\Gamma_{21}, \Gamma_{22} \in \mathbb{R}^{m\times n} $ are full-rank matrices, and $\Gamma_{33} \in \Rr^{m \times m}$ is positive definite.
To use the contraction property mentioned in Theorem \ref{Th1}, we recast the closed-loop system \eqref{Eq:hamiltonian3-d} with new energy function $\bar H_{d_3}(q,p,\mathcal{Z},t)$ as follows\footnote{We omit the argument $t$ in $z_1$, $z_2$, and $\gamma_2$.}
\begin{equation}
\label{Eq:target3}
\begin{bmatrix}
{{\dot{q}}} \\[.05cm]
{{\dot{p}}}\\[.05cm]
{{\dot{\mathcal{Z}}}}
\end{bmatrix}=\begin{bmatrix}
0 & J_{d_{12}} & 0\\[.05cm]
-J_{d_{12}}^{\top} & 0&\digamma_1\\[.05cm]
\digamma_2 & 0 & \digamma_3\\[.05cm]                    
\end{bmatrix}\begin{bmatrix}
{{\nabla_{{q}}}{\bar{H}_{d_3}(q,p,\mathcal{Z},t)}}\\[.05cm]
{{\nabla_{{p}}}{\bar{H}_{d_3}(q,p,\mathcal{Z},t)}}\\[.05cm]
{{\nabla_\mathcal{Z}}{\bar{H}_{d_3}(q,p,\mathcal{Z},t)}}\\[.05cm]         \end{bmatrix},
\end{equation} 
\begin{equation}
\arraycolsep=1pt
\def\arraystretch{1.2}
\begin{array}{rcl}
\label{Eq:hamiltonia3}
&\bar{H}_{d_3}(q,p,\mathcal{Z},t)=\frac{1}{2}p^\top M_d^{-1} p
+V_{d_3}(q,z_1,t)\\&+\frac{1}{2}(z_2-\gamma_2+\mu_2)^{\top}K_{z}(z_2-\gamma_2+\mu_2),
\end{array}
\end{equation} 
where $\mu_2 \triangleq  (\Gamma_{12} K_z)^\dagger  Gd$. 

Derived from Theorem \ref{Th1}, the following corollary establishes conditions to guarantee that \eqref{Eq:target3} is a contractive system using similar arguments as the ones given in Corollaries \ref{Cor:contract-w} and \ref{Cor:contract-r}.
\begin{Cor}
	\label{Cor:contract-w-r}
	Consider the closed-loop dynamics \eqref{Eq:target3}. Suppose that:   
	\begin{itemize}
		\item [\textbf{C1.}] The matrix
		\begin{equation}
		\label{B_d}
		\begin{split}
		P_3 \triangleq \begin{bmatrix}
		0 & J_{d_{12}} & 0\\[.05cm]
		-J_{d_{12}}^{\top} & 0&\digamma_1\\[.05cm]
		\digamma_2 & 0 & \digamma_3\\[.05cm]                    
		\end{bmatrix},
		\end{split}
		\end{equation} 
		is Hurwitz.
		\item [\textbf{C2.}] There exist constants $0\!<\alpha_3<\beta_3$ such that $ \bar H_{d_3}(\theta,t)$ in \eqref{Eq:hamiltonia3}, with $\theta=[q^\top,p^{\top},\mathcal{Z}^\top]^\top$, satisfies the inequality
		\begin{equation}
		\label{Eq:bondcond3}
		\alpha_3 I\prec \nabla_{\theta}^2 \bar H_{d_3}(\theta,t) \prec \beta_3 I,\; \forall \theta \in D_3\subseteq \Rr^{2n+2m},
		\end{equation}
	\end{itemize}

	The system $\eqref{Eq:target3}$ is contractive on $D_3$ if there exists a positive constant  $\varepsilon_3$ such that 	
	\begin{equation}
	\label{N3}
	N_3 = \begin{bmatrix}
	P_3 & (1-\frac{\alpha_3}{\beta_3}) P_3P_3^\top\\ -(1-\frac{\alpha_3}{\beta_3}+\varepsilon_3)I & -P_3^\top
	\end{bmatrix},
	\end{equation}	
	has no eigenvalues on the imaginary axis.   $\square$
\end{Cor}

Now, we propose a robust tracking controller without velocity terms.  To shorten the notation in the rest of the paper, we define $\Phi(q,z_1,t)\triangleq  \nabla_q V_{d_3}(q,z_1,t)$.
 \begin{Thm}
	\label{Pro:UT-r-w}
	Consider the mechanical system \eqref{open-loop mech} under Assumptions \ref{ass:M} and \ref{ass:G}, with a feasible trajectory $x^\star(t)\!=\![{q^\star}\!^\top\!\!,{p^\star}\!^\top]^\top\!$. 
	If there exist ${J_d}_{12}$, $M_{d}$, $V_{d_3}$, $K_z$ and $\digamma_i$, for $i\kern-.3em\in\kern-.3em \{1,2,3\}$, in \eqref{Eq:digamma12} such that:
	
 \begin{enumerate}[label=(\roman*)]
		\item The following matching equations hold
  
		\begin{eqnarray}
		&&\hspace{-1.2 cm}M^{-1}p=J_{d_{12}}M_d^{-1}p, \label{Eq:matcing3-p1}\\
		&&\hspace{-1.2 cm}G^\perp\Big(\nabla_q V(q)-J_{d_{12}}^{\top}\nabla_q V_{d_3}(q,z_1,t)\nonumber\\&&\hspace{-1.2 cm}+\Gamma_{11}\nabla_{z_1}V_{d_3}(q,z_1,t)+\Gamma_{12}K_{z}(z_2-\gamma_2)\Big)=0\label{Eq:matcing3-p2}
		\end{eqnarray}

		\item The conditions mentioned in Corollary \ref{Cor:contract-w-r} are satisfied. 	
		\item The following equations are satisfied
			\begin{align}
			&{{\dot z_1}^{\star}}=\Gamma_{21}\Phi(q^{\star} ,z_1^{\star},t)-\Gamma_{33}\nabla_{z_{1}}V_{d_3}(q^{\star},z_1^{\star},t)\nonumber\\&
		    0=\Gamma_{22}\Phi(q^{\star}, z_1^{\star},t)\nonumber
			\end{align}
	\end{enumerate}
    where ${p^\star}= M\dot q^\star$.
	The controller 
	\begin{equation}
	\label{Eq:UT-w-r}
	\begin{split}
	u =G^ \dagger&\Big(-J^\top_{d_{12}} {\Phi}(q,z_1,t)+\Gamma_{11}\nabla_{z_1}{{V_{d_3}}(q,z_1,t)}\\&+\Gamma_{12} K_z (z_2-\gamma_2)+
	\nabla_q H\Big),
	\end{split}
	\end{equation}
	with
	\vspace{-0.2cm}
	\begin{equation}
	\label{Eq:gamma-r-w}
\begin{array}{cl} &\gamma_2(t)=(G^\top\Gamma_{12}K_z)^{-1}\big(G^\top (-J^\top_{d_{12}} {\Phi}(q^\star,z^\star,t)\\& +G^\top\Gamma_{11}\nabla_{z_1}{{V_{d_3}}(q^\star,z_1^\star,t)}-G^\top \dot{p}^\star\big),
	\end{array} 
	\end{equation}
	and
    \vspace{-0.3cm}
\begin{equation}
\label{Eq:extend3}
\begin{array}{cl}
&{{\dot z_1}}=\Gamma_{21}\Phi(q,z_1,t)-\Gamma_{33}\nabla_{{z_1}}V_{d_3}(q,z_1,t)\\&
{{\dot z_2}}=\Gamma_{22}\Phi(q,z_1,t)
\end{array}
\end{equation}
	realizes exponential tracking of $x^\star(t)$ without velocity measurements while eliminating the disturbance effect.
	\begin{pf}
Because of (i), substituting \eqref{Eq:UT-w-r} into \eqref{open-loop mech}  yields \eqref{Eq:target3}. Moreover, (ii) ensures that the closed-loop system \eqref{Eq:target3} is contractive.
We next evaluate the desired signals $(x^\star,z_1^\star, z_2^\star)$, where $z_2^\star \in {\mathbb{R}^{m}}$ is constant, in the contractive system \eqref{Eq:target3} as follows
		\begin{align}
        &\dot p^\star=-J^\top_{d_{12}}\nabla_q \bar H_{d_3}(\theta^\star,t)\label{Eq:ref1-r-w}\\&+\Gamma_{11}\nabla_{z_1}V_{d_3}(q^\star,z_1^\star,t)+\Gamma_{12}K_{z}(z^\star_2-\gamma_2+\mu_2),\nonumber
         		\end{align}
		\begin{align}
        &{{\dot z_1}^{\star}}=\Gamma_{21}\nabla_q \bar H_{d_3}(\theta^\star,t)-\Gamma_{33}\nabla_{z_{1}}V_{d_3}(q^{\star},z_1^{\star},t)\label{Eq:ref2-r-w}\\&
       0=\Gamma_{22}\nabla_q \bar H_{d_3}(\theta^\star,t)\label{Eq:ref3-r-w}
		\end{align}
		
		Then, we multiply both sides of \eqref{Eq:ref1-r-w} by the invertible matrix $\small[G^\perp,G^\top]^\top$ and replace $\nabla_q \bar H_{d_3}(\theta^\star,t)\!=\!\Phi(q^{\star},z_{1}^{\star},t)$ in \eqref{Eq:ref1-r-w}-\eqref{Eq:ref3-r-w}. Since $(\mbox{iii})$ and \eqref{Eq:gamma-r-w}, we conclude that \eqref{Eq:ref1-r-w}--\eqref{Eq:ref3-r-w} are satisfied with $z_2^\star\!=\!-\mu_2$. Hence, convergence property in the contractive system implies that all the trajectories of \eqref{Eq:target3} exponentially converge to the desired ones. $\blacksquare$ 
	\end{pf}
\end{Thm}

Subsections \ref{sec:full} and \ref{sec:under} study how to apply the result of Theorem \ref{Pro:UT-r-w} to two classes of mechanical systems.

\subsection{Fully actuated mechanical systems}\label{sec:full}
To introduce the result of this subsection, we consider that $V_{d_3}(q,z_1,t)$ is given by
\begin{equation}
\label{Eq:potential-fully}
\arraycolsep=0.5pt
\def\arraystretch{1}
\begin{array}{rcl}
V_{d_3}(q,z_1,t)&=&\frac{1}{2}(q-L(t))^\top K_q(q-L(t))\\&&+\frac{1}{2}(q-z_1)^\top K_c(q-z_1),
\end{array}
\vspace{-0.1cm}
\end{equation}
where $K_q,K_c \in \Rr^{n \times n}$ are positive definite and $L\!:\!\Rr_{+}\to\Rr^{n}$ is given by
\begin{equation}
    L(t)=K_q^{-1}K_c(q^\star-z^\star_1)+q^\star.\label{Eq:L3}
\end{equation}
Hence,
\begin{align}
&
\nabla_{{z_1}}V_{d_3}(q,z_1,t)=-K_c(q-z_1),\label{Eq:gradVd3}\\&
\Phi(q,z_1,t)=K_q(q-L(t))+ K_c(q-z_1)\nonumber.
\end{align}

The next proposition establishes a set of criteria, based on Theorem \ref{Pro:UT-r-w}, to design a controller for fully actuated mechanical systems. 

\begin{pro}
\label{Cor:fully actuated} 
Consider \eqref{open-loop mech} with $n=m$, $G=I$, constant $M$ , and a  feasible trajectory $x^\star(t)\!=\![{q^\star}\!^\top\!\!,{p^\star}\!^\top]^\top\!$. Set $M_d=M$, $J_{d_{12}}\!=\!I$, and $V_{d_3}(q,z_1,t)$ as in \eqref{Eq:potential-fully}. 
Select the parameters $K_q, K_c, K_z$, $\digamma_1$, $\digamma_2$, and $\digamma_3$ such that the conditions in Corollary \ref{Cor:contract-w-r} are satisfied. The input signal 
\begin{equation}
\begin{array}{rcl}
\label{Eq:UT-w-f}
u\!&=\!\!-K_q(q-L(t))-(I+\Gamma_{11})K_c(q-z_1)\nonumber\\&+\Gamma_{12} K_z (z_2-\gamma_2)+\nabla_q V(q),
\end{array}
\end{equation}
with $L(t)$ given by \eqref{Eq:L3} and
\begin{align}
&\dot z^\star_1=\Gamma_{33} K_c (q^\star-z^\star_1)\label{Eq:z-1},\\
&\gamma_2=(\Gamma_{12}K_z)^{-1}\big(-\Gamma_{11}K_c(q^\star-z^\star_1)-\dot p^\star\big),\label{Eq:gamma-R}
\end{align}
and
\begin{align}
&\dot z_1=\Gamma_{21}K_q(q-L)\nonumber+(\Gamma_{21}+\Gamma_{33})K_c(q-z_1),\label{Eq:extend3-f1}\\&
\dot z_2=\Gamma_{22}\big(K_q(q-L)+K_c(q-z_1)\big)\nonumber,
\end{align}
realizes exponential robust tracking of $x^\star(t)$ without requiring velocity measurements.
\begin{pf}
We want to prove that the conditions in Theorem \ref{Pro:UT-r-w} are satisfied. To this end, note that $G^\perp=0$. Hence, the matching equations \eqref{Eq:matcing3-p1}-\eqref{Eq:matcing3-p2}, are (trivially) satisfied. Furthermore, the conditions in Corollary \ref{Cor:contract-w-r} are met by construction. Consequently, (ii) in Theorem \ref{Pro:UT-r-w} holds.
From \eqref{Eq:L3}, \eqref{Eq:gradVd3} and \eqref{Eq:z-1} evaluated at $(q^\star,p^\star,t)$, it follows that $(\mbox{iii})$ in Theorem  \ref{Pro:UT-r-w} holds. $\blacksquare$	
\end{pf}			
\end{pro}
\vspace{-1 mm}
\subsection{Underactuated mechanical systems}\label{sec:under}
To establish the result of this subsection, we consider
\begin{equation}
\label{Eq:potential-under}
\arraycolsep=1pt
\def\arraystretch{1.2}
\begin{array}{rcl}
V_{d_3}(q,z_1,t)&=& \phi_1(q)+\frac{1}{2}k_1(\phi_2(q)-\ell_3(t))^2\\&&+\frac{1}{2}k_2(\phi_2(q)-\phi_3(z_1))^2,
\end{array}
\end{equation}
where $k_1\!>\!0, k_2\!>\!0$, $\phi_1, \phi_2\!:\!\Rr^n\!\to \Rr$, $\phi_3\!:\!\Rr^m\! \to \Rr$ and $\ell_3: \Rr_{+} \to \Rr $ is given by
\begin{equation}
\begin{array}{l}
\ell_3(t)=\left(\frac{(\Gamma_{22}\nabla_q\phi_2)^\dagger }{k_1}\Gamma_{22}\right)\big(\nabla_q\phi_1(q^\star)\\+k_1\phi_2(q^\star)\nabla_q\phi_2(q^\star)
+k_2(\phi_2(q^\star)-\phi_3(z^{\star}_1))\nabla_q\phi_2(q^\star)\big),
\end{array}
\label{Eq:l-R-W}
\end{equation}
Thus, 
\begin{equation}
\label{Eq:Vd}
\begin{split}
&\nabla_{z_1} V_{d_3}(q,z_1,t)=-k_2(\phi_2(q)-\phi_3(z_1))\nabla_{z_1}\phi_3(z_1),\\&
\Phi(q,z_1,t)=\nabla_q\phi_1(q)+k_1(\phi_2(q)-l_3(t))\nabla_q\phi_2(q)
\\&+k_2(\phi_2(q)-\phi_3(z_1))\nabla_q\phi_2(q).
\end{split}
\end{equation}

Based on Theorem \ref{Pro:UT-r-w}, the following proposition establishes conditions to design robust---with respect to constant matched disturbances---controllers that achieve trajectory tracking for underactuated mechanical systems without requiring velocity measurements. 
\begin{pro}
\label{Cor:underactuated}	
Consider \eqref{open-loop mech}, with $m<n$,  constant $M$, constant $G$, and a feasible trajectory $x^\star(t)\!=\![{q^\star}\!^\top\!\!,{p^\star}\!^\top]^\top\!$. Consider $V_{d_3}(q,z_1,t)$ as in \eqref{Eq:potential-under} with 
$l_3(t)$ given by \eqref{Eq:l-R-W} and
\begin{align}
&
\dot z^\star_1(t)=\Gamma_{33}k_2\big(\phi_2(q)-\phi_3(z_1)\big)\nabla_{z_1}\phi_3(z_1).\label{Eq:z-R-W}
\end{align}
Consider one of the following cases:	
\begin{enumerate}
	\item [(a)] Only potential energy shaping. Hence, $M_d=M$, $J_{d_{12}}\!=\!I$.	
	\item [(b)] Total energy shaping. Hence, $J_{d_{12}}\!=\!M^{-1}M_{d}$.
\end{enumerate}	
If the parameters $k_1, k_2\!\!>\!\!0, K_z\!\!\succ\!\!0, \digamma_i$, $\phi_i$, for $i\in \{1,2,3\}$, satisfy the conditions in Corollary \ref{Cor:contract-w-r} and \eqref{Eq:matcing3-p2},
then it follows from \eqref{Eq:Vd} that the control law \eqref{Eq:UT-w-r}, with the desired potential energy given in  \eqref{Eq:potential-under}, guarantees that the closed-loop system tracks $x^\star(t)$ and is robust with respect to constant matched disturbances.
\begin{pf}
 The matching equations \eqref{Eq:matcing3-p1} are satisfied in scenarios (a) and (b).  Besides, the conditions in Corollary \ref{Cor:contract-w-r} and \eqref{Eq:matcing3-p2} are satisfied by construction. Hence, (i) and (ii) in Theorem \ref{Pro:UT-r-w} hold. Given \eqref{Eq:l-R-W}, \eqref{Eq:Vd}, \eqref{Eq:z-R-W} and the corresponding $\Phi(q,z_1,t)$, (iii) in Theorem \ref{Pro:UT-r-w} is satisfied. $\blacksquare$
\end{pf}	    
\end{pro}	
\begin{rem}
\label{solvability3}
Following a similar rationale as the one in Section \ref{se:Par-cte}, suppose $\digamma_1=GK_f$, where $K_f \in \mathbb{R}^{m \times 2m}$. The matching equation \eqref{Eq:matcing3-p2} is reduced to 
\begin{equation}
\arraycolsep=1pt
\def\arraystretch{1}
\begin{array}{rcl}
\label{Eq:matcing1-p31}
&&G^\perp\Big(\nabla_q V(q)-J_{d_{12}}^{\top}\nabla_q V_{d_3}(q,z_1,t)\Big)=0.
\end{array}
\end{equation}

Note that the set of solutions to \eqref{Eq:matcing1-p31} is the same as the set of solutions to \eqref{solvability22} in the conventional IDA-PBC problem.
\end{rem}
\begin{rem}
If the feasible trajectory $x^\star(t)$ is chosen constant in Theorems \ref{Pro:UT-w}, \ref{Pro:UT-r} and \ref{Pro:UT-r-w}, then the proposed controllers solve the regulation problem while guaranteeing exponential stability of the desired equilibrium $x^\star$.  
\end{rem}
\vspace{-0.2 cm}
\section{Simulation}
\label{sec:sim}
In this section, we illustrate the effectiveness of the results proposed in Section \ref{sec:no-velocity-robust}. To this end, we solve the trajectory-tracking problem for an underactuated mechanical system, namely, the ball-on-wheel system (see Fig. \ref{ball-wheel}). 
 
The dynamics of the ball on wheel system with constant matched disturbances are given by \eqref{open-loop mech}  with input matrix $G=[0,1]^\top$ and the state and input dimensions $n=2$ and $m=1$, respectively. The system states $q_1$ and $q_2$ are the angular displacement of the contact point between the ball and the wheel and the angular displacement of the wheel, respectively. The Hamiltonian potential energy and the inertia matrix are given by
\begin{equation*}
\begin{split}
&V(q)=m_4\cos(q_1),\quad m_4=m_bg_r(r_w+r_b),\\&
M\!=\!\begin{bmatrix}
m_1 & m_2\\
m_2 & m_3
\end{bmatrix}\!=\!\begin{bmatrix}
\bigg(\frac{2}{5}+m_b\bigg)(r_w+r_b)^2 & -\frac{2}{5}({r_w}^2+r_wr_b)\\
-\frac{2}{5}({r_w}^2+r_wr_b) & I_w+\frac{2}{5}r_w^2
\end{bmatrix},
\end{split}
\end{equation*}
where $m_b,r_w,r_b,g_r$, and $I_w$ are the mass of the ball, the radius of the wheel, the radius of the ball, the gravity acceleration
and the moment inertia of the wheel, respectively. 

The closed-loop mechanical system is of the form \eqref{Eq:target3} and \eqref{Eq:hamiltonia3}.  To achieve the total energy shaping objective in Corollary \ref{Cor:underactuated}, $M_d$ is characterized as follows
\begin{equation*}
M_d= \left[ {\begin{array}{*{20}{c}}
	a_1&a_2 \\ 
	a_2&a_4 
	\end{array}} \right], \quad a_1,a_3 >0,\,\, a_1a_3>a^2_2,
\end{equation*}
$\digamma_1$ is chosen according to Remark \ref{solvability3}. Therefore, the matching equation \eqref{Eq:matcing1-p3} is simplified as
\begin{equation}
\label{Eq:matcing1-ball}
\begin{split}
[1,0]&\big(-m_4\sin(q_1)-M_dM^{-1}\nabla_q V_d(q,z_1,t)\big)=0.
\end{split}
\end{equation}

Then, the solution to \eqref{Eq:matcing1-ball} is determined as the potential energy function \eqref{Eq:potential-under} with the following ingredients
\begin{equation*}
\begin{split}
&\phi_1(q)=\lambda_1cos(q_1),\quad
\phi_2(q)=\lambda_2 q_1+q_2 ,\quad
\phi_3(z_1)=z_1,
\end{split}
\end{equation*} 
where
\begin{equation*}
\begin{split}
&\lambda_1=\frac{m_4(m_1m_3-m_2^2)}{a_1m_3-a_2m_2},\quad
\lambda_2=\frac{m_2a_1-m_1a_2}{a_1m_3-a_2m_2}.\quad
\end{split}
\end{equation*}

Note that the solution to \eqref{Eq:matcing1-ball} can be determined based on the general solution to the matching equation in \cite[Chapter~6]{yaghmaei2019output}, where the timed IDA-PBC approach is investigated for this system.

\begin{figure}
	\centering
	\includegraphics[width=0.3\columnwidth]{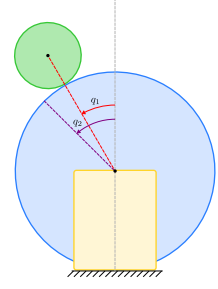}
	\centering
	\vspace{0.1cm}
	\caption{Schematic of the ball on wheel system}
	\label{ballonwheel-eps-converted-to}
\end{figure}
\begin{figure}
	\centering
	\includegraphics[width=0.9\columnwidth]{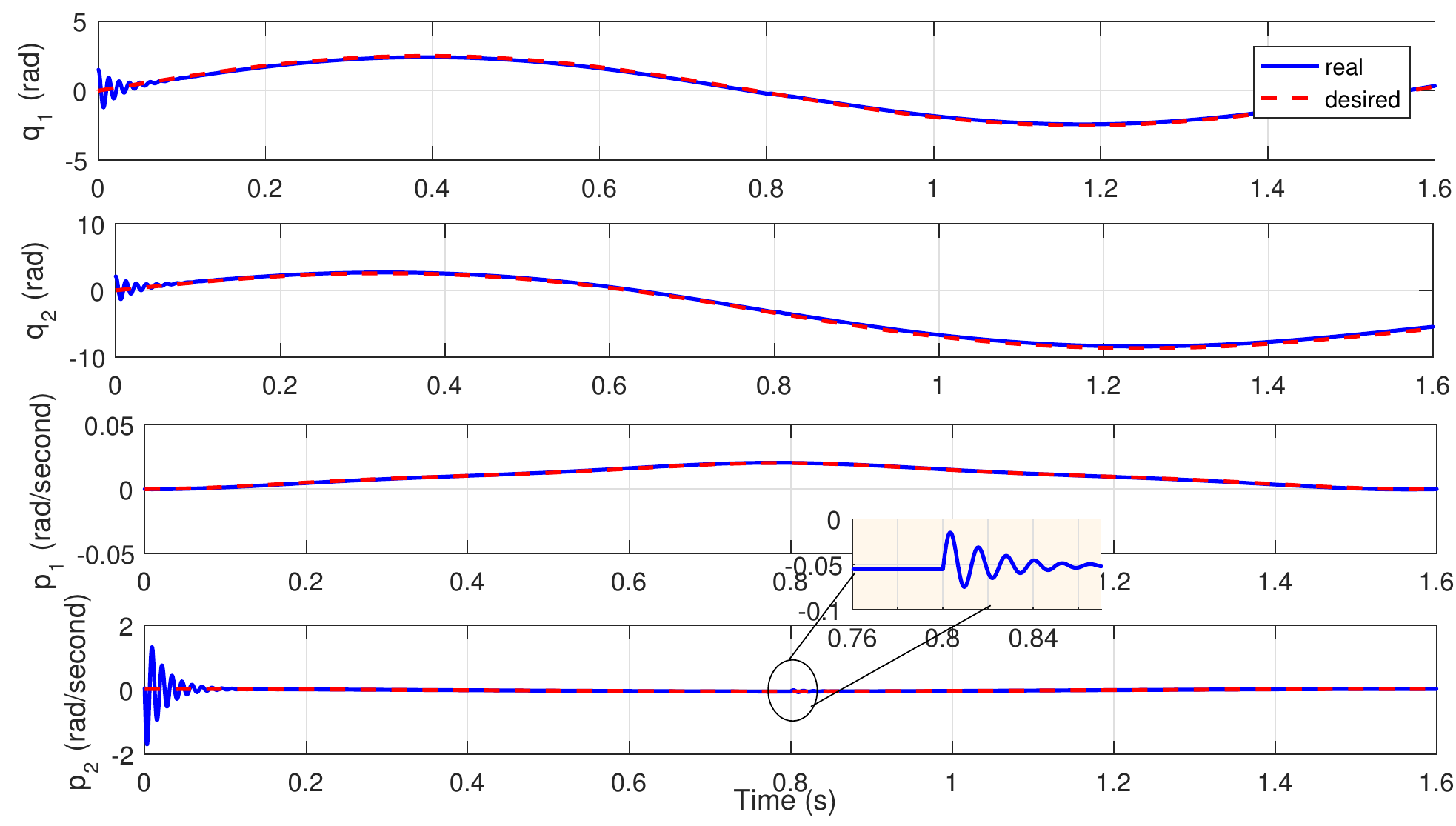}
	\centering
	\vspace{-0.3cm}
	\caption{ \small Desired and closed-loop trajectories for the ball-on-wheel system. Note that the angular displacement $q_1$ exponentially tracks the desired signal $a(t)=2.5\sin(4t)$.}
	\label{ball-RW}
\end{figure}
Now, by selecting suitable values of the parameters $k_1, k_2>0$, $K_z \succ0$, $\digamma_i$, for $i\in\{1,2,3\}$, the conditions in Corollary \ref{Cor:contract-w-r} are satisfied. Thereby, the robust tracking controller without velocity terms \eqref{Eq:UT-w-r} using the trajectory gains  \eqref{Eq:gamma-r-w} and \eqref{Eq:l-R-W} is designed to track the reference $x^\star(t)$ with the matched constant disturbance $d$, which is  added to the system at $t=0.8s$. We use the numerical values stated in Table \ref{tab:1} for simulation purposes. The results are depicted in Fig. \ref{ball-RW}. Note that the angular displacement of the contact point (i.e, $q_1(t)$) exponentially tracks the desired signal $a(t)=2.5\sin(4t)$, while the effect of the disturbance $d$ is eliminated. Besides, the desired trajectory can be computed based on Definition \eqref{definition1} as follows
\begin{equation*}
\begin{split}
&q^\star_1(t)=a(t),\quad p^\star_1(t)=\int_{o}^{t} m_4\sin(a(\tau))d_\tau+b_0,\\&
p^\star_2(t)=\frac{m_3}{m_2}\int_{o}^{t}m_4\sin(a(\tau))d_\tau+\frac{m_3}{m_2}b_0-\frac{m_1m_3-m^2_2}{m_2}\dot a(t),\\&
q^\star_2(t)=\frac{1}{m_2}\int_{o}^{t}\int_{o}^{\tau} m_4\sin(a(\sigma))d_\sigma d_\tau-\frac{m_1}{m_2}a(t)+b_1,\\&
u^\star(t)=\frac{m_3}{m_2} m_4\sin(a(t))-\frac{m_3}{m_2}b_0-\frac{m_1m_3-m^2_2}{m_2}\ddot{a}(t).
\end{split}
\end{equation*}
\setlength{\arrayrulewidth}{.5mm}
\setlength{\tabcolsep}{1.5pt}
\renewcommand{\arraystretch}{1.}
\begin{table}[t!]
	\scriptsize
	\centering
	\caption{\small Control Parameters.}
	\vspace{-0.4cm}
	\begin{tabular}{ p{0.09\textwidth} c c c }
		\\[-1ex]\hline\\[-1ex]
		\textbf{Open-loop}&$I_w=0.00171$&$m_b=0.042$&$r_b=0.011$\\&$g_r=9.8$&$r_w=0.075$\\[1ex] 
		\hline\\[-1ex]
		\textbf{Closed-loop}
		&$a_1= 4 \times 10^{-3}$&$ a_2=-4.8 \times 10^{-3}$& $a_3=0.04$\\[1ex]  
		&$k_1=1.8$&$k_2=3.5$&$ K_{z}=0.1163$\\[1ex]
		&$\Gamma_{11}^\top=[0,5]$&$\Gamma_{12}^\top=[0,0.6]$&$\Gamma_{33}=26.8$\\[1ex] &$\Gamma_{21}=[5,0]$&$\Gamma_{22}=[-0.005,0]$&$d=20$\\  [1ex] 
		\hline
	\end{tabular}
	\label{tab:1}
\end{table} 
\vspace{-6pt}
\section{Conclusion}
\label{sec:con}
\vspace{-4pt}
This paper proposes an approach to address the tracking problem without velocity measurements for mechanical systems  ---fully actuated and underactuated cases---  subject to matched constant disturbances. To this aim, we use the contraction property of the desired system interconnected to the dynamic extension. The proposed method utilizes an extended form of the IDA-PBC technique.
 The suggested controller shows positive results in the simulation of the ball on a wheel system.
\begin{ack}
The authors thank Jos\'{e} \'{A}ngel Acosta for his feedback on the previous version of this paper.
\end{ack}

\bibliography{ifacconf}             
                                                   







\end{document}